\documentclass[twocolumn,nofootinbib,pra,showpacs,preprintnumbers,amsmath,amssymb,a4paper]{revtex4-1}
\usepackage{hyperref} 
\usepackage{graphicx}%
\usepackage{url}
\hypersetup{colorlinks,breaklinks,
citecolor=[rgb]{0.0,0.5,0.5},
    urlcolor=[rgb]{0.0,0.5,0.5},
    linkcolor=[rgb]{0.0,0.5,0.5}}
 
\newtheorem{theorem}{Theorem}[section]       

\DeclareMathOperator{\Tr}{Tr}

\def\ket #1{\vert #1\rangle}
\def\bra #1{\langle #1\vert}
\newcommand{\ketbra}[2]{\ensuremath{\ket{#1}\!\bra{#2}}}
\newcommand{\braket}[2]{\ensuremath{\bra{#1}{#2}\rangle}}
\newcommand{\Jam}{Jamio\l kowski }
\newcommand{\Zd}{\ensuremath{\mathbb{Z}_d}}
\newcommand{\Zp}{\ensuremath{\mathbb{Z}_p}}
\newcommand{\Fq}{\ensuremath{\mathbb{F}_q}}
\newcommand{\Fp}{\ensuremath{\mathbb{F}_p}}
\DeclareMathOperator*{\argmin}{argmin}

\usepackage[normalem]{ulem}

\begin{document}

\title{Maximum nonlocality and minimum uncertainty using magic states}
\author{Mark Howard}
\affiliation{%
Institute for Quantum Computing and Department of Applied Mathematics,
University of Waterloo, Waterloo, Ontario, Canada, N2L 3G1
}
\email{mark.howard@uwaterloo.ca}

\begin{abstract}

We prove that magic states from the Clifford hierarchy give optimal solutions for  tasks involving nonlocality and entropic uncertainty with respect to Pauli measurements.
 For both the nonlocality and uncertainty tasks, stabilizer states are the worst possible pure states so our solutions have an operational interpretation as being highly non-stabilizer. The optimal strategy for a qudit version of the Clauser-Horne-Shimony-Holt (CHSH) game in prime dimensions is achieved by measuring maximally entangled states that are isomorphic to single-qudit magic states. These magic states have an appealingly simple form and our proof shows that they are ``balanced'' with respect to all but one of the mutually unbiased stabilizer bases. Of all equatorial qudit states, magic states minimize the average entropic uncertainties for collision entropy and also, for small prime dimensions, min-entropy -- a fact that may have implications for cryptography.

\end{abstract}

\maketitle

\section{Introduction and Motivation}

In the context of fault-tolerant quantum computing using error-correcting codes, there exist single-particle pure states --``magic states'' -- with desirable properties. Having access to a supply of these states expands the sets of robustly implementable operations to encompass a universal gate set. Moreover, impure versions of these states can be purified using only the fault-tolerant operations provided by the error-correcting code. There is a privileged family of magic states, well-defined for both qubits and qudits of odd prime dimension $p$, whose structure reflects a group-theoretical object called the Clifford hierarchy. Here we will show that the utility of these magic states extends beyond the arena of quantum computation; they essentially provide optimal strategies for both (i) a nonlocal game wherein both Alice and Bob each use $p$ Pauli measurements, and (ii) a cryptographically motivated scenario wherein we try to minimize the average entropic uncertainty with respect to a set of $p$ mutually unbiased measurements. Qudits can have operational advantages for nonlocal or cryptographic tasks in terms of tolerating inefficiencies or the addition of white noise \cite{Durt:2003,Acin:2002,Acin:2004}, but the geometry of state space becomes notoriously complex when we venture past the qubit Bloch sphere. In this work we prove that the optimal states have a simple structure and furthermore that these states have connections to group-theoretical and number-theoretical structures. These ideas may prove useful elsewhere.

Given the fundamental importance of the Clauser-Horne-Shimony-Holt \cite{CHSH:1969} (CHSH) game in quantum information theory \cite{Brunner:2014,Buhrman:2010,Reichardt:2013}, it is well-motivated to study generalizations to different numbers of measurement settings, parties and so on. By preserving the structure of the CHSH game while enlarging the size of the input/output alphabet to $d$ symbols (using qudits), we can examine how the quantum advantage scales with dimension or investigate new features that arise due to additional degrees of freedom. Buhrman and Massar \cite{Buhrman:2005} studied such a generalization and found bounds on the allowable quantum value. Ji \emph{et al.} \cite{Ji:2008} constructed a Bell operator from qudit Pauli measurements, whose maximization describes a quantum strategy for the qudit CHSH game. For small prime dimensions they analyzed classical (i.e. local hidden variable) and quantum values. A follow-up work by Liang \emph{et al.} \cite{Liang:2009} expanded on this analysis, and broadened the numerical search to allow non-Pauli measurements. Most recently, Bavarian and Shor \cite{Bavarian:2013} looked at a generalized CHSH game wherein the alphabet consists of all field elements $\Fq$ where $q$ is a prime power, and proved a generalization of Tsirelon's bound using Information Causality \cite{Pawlowski:2009}. As was done by Ji \emph{et al.} \cite{Ji:2008}, we will restrict our analysis to projective Pauli measurements. Quantifying nonlocality is a tricky task, particularly if one is motivated by operational considerations. For example, an experimental test may achieve higher statistical significance via lower statistical error  \cite{Jungnitsch:2010} or higher statistical strength via larger statistical divergence with any classical model \cite{Acin:2005}. One conceptually simple metric for nonlocality is to measure the amount by which a Bell inequality is violated. If Bell inequality violation is the only figure of merit then restricting to Pauli measurements is suboptimal; larger violations are possible with non-Pauli measurements whenever $p>3$. Nevertheless, we remain particularly interested in non-stabilizer states, measurements (see e.g \cite{Renes:2004}) or transformations that emerge as privileged with respect to their stabilizer counterparts, due to some geometrical or operational relationship. This is motivated in part by the clean mathematical and conceptual framework provided by the so-called ``stabilizer subtheory''\cite{Veitch:2012,Gross:2006,Veitch:2014} of quantum mechanics for odd-dimensional qudits -- loosely speaking, states and operations far outside this subtheory are highly non-classical \cite{Howard:2014}.

Mutually unbiased bases (MUBs) are a crucial component in a variety of quantum-informational settings. They are necessary for the creation of maximally strong entropic uncertainty relations \cite{Wehner:2010}, which describe limits on the allowed probability distributions associated with multiple measurements on quantum states. In cryptographic scenarios, uncertainty relations can be used to bound an eavesdropper's knowledge of quantum states transmitted between two parties. In \cite{Amburg:2014} Amburg \emph{et al.} motivate the study of ``MUB-balanced states'' i.e., pure states for which, relative to a complete set of $d+1$ MUBs, the list of probabilities of the $d$ outcomes of one of these measurements is independent of the choice of measurement, up to permutations. Recent work by Appleby \emph{et al.} \cite{Appleby:2014} suggests that MUB-balanced states only exist in odd prime-power dimensions of the form $d=3 \bmod 4$. A MUB-balanced state is automatically also a collision-entropic minimum uncertainty \cite{Appleby:2014,Wootters:2007} (or maximally certain \cite{Mandayam:2010}) state, important in cryptography, but balancedness is a stricter condition than minimum uncertainty. From a foundational point of view, MUB-balanced states are analagous to harmonic oscillator eigenstates insofar as the ``direction'' of a measurement is unimportant. Our results include a concise proof that magic states are balanced with respect to $p$ out of a total of $p+1$ total bases in a complete set of MUBs. This suggests that magic states may be relevant in quantum cryptography, in analogy with the Breidbart basis for qubits. We also comment on the connection between magic states and the Sato-Tate distribution that arises in number theory.

\section{Background Material}
\subsection{Prerequisites and Notation}\label{sec:Prerequisites}
The Weyl-Heisenberg operators $D_{(x|z)}$ are a generalized $p$-dimensional version of qubit Pauli operators
\begin{align}
&D_{(x|z)}=\omega^{2^{-1}xz}\sum_k \omega^{kz}\ketbra{k+x}{k}=\omega^{\frac{xz}{2}}X^xZ^z \label{eqn:DispOps}\\
&\text{with }\quad X\ket{j}=\ket{j+1} \quad Z\ket{j}=\omega^j\ket{j} \quad \left(\omega=e^{2\pi i/p}\right),\nonumber
\end{align}
where we always treat expressions like $\frac{1}{a}$ to mean the mutiplicative inverse $a^{-1} \in \Zp=\{0,1,\ldots,p-1\}$.
The choice of phase has the convenient feature that $D_{(x|z)}^n=D_{(nx|nz)}$ and we use symplectic notation to keep track of $X$ and $Z$ powers separately, so that multi-qudit operators look like
\begin{align*}
&D_{(x_1|z_1)} \otimes D_{(x_2|z_2)} =D_{(x_1,x_2|z_1,z_2)}.
\end{align*}

The rank-1 projector onto the $\omega^V$ eigenspace of $D_{(1|B)}$ (in other words, the projector onto the $V$-th vector of the $B$-th Weyl-Heisenberg basis) is given by
\begin{align}
&\ketbra{\psi_B^V}{\psi_B^V}=\frac{1}{p}\sum_j \omega^{-jV}D_{(1|B)}^j \label{eqn:projBV},\\
\text{so that }&\ket{\psi_B^V}=\frac{1}{\sqrt{p}}\sum_{k \in \mathbb{F}_q} \omega^{(\frac{1}{2}Bk^2-Vk)}\ket{k}\label{eqn:ketBV}.
\end{align}
 The $p^2$ states $\{\ket{\psi_B^V}, B,V \in \Zp\}$, along with those of the computational basis, comprise both (i) a complete set of mutually unbiased bases (ii) the complete set of stabilizer states (Pauli eigenstates), for a single qudit.

Consider the family of magic states $f$ and magic gates $M$ (diagonal in the computational basis) \cite{Howard:2012,Campbell:2012,Campbell:2014}
\begin{align}
\ket{f_{a,b,c}}=&\frac{1}{\sqrt{p}}\sum_k \omega^{ak^3+bk^2+ck}\ket{k} \quad p> 3 \label{eqn:fabcdef} \\
&=M_{a,b,c}\ket{+}\qquad a\in \Zp^*,\,b,c\in \Zp 
\end{align}
where $\ket{+}$ is the equal-weighted superposition of all computational basis states and 
$\Zp^*=\{1,2,\dots,p-1\}$. 
Note that the above expressions require a minor modification for the smallest prime dimensions \cite{Howard:2012} i.e.,
\begin{align}
\ket{f_{a,b,c}}=\begin{cases} \frac{1}{\sqrt{2}}(\ket{0}+\gamma^{a+2b+4c}\ket{1}) \qquad &p=2
\\
\frac{1}{\sqrt{3}}(\ket{0}+\xi^{2a+6b+3c}\ket{1}+\xi^{a+6b+6c}\ket{2}) 
\quad &p=3\end{cases} \label{eqn:fabc23}
\end{align}
where $\gamma=e^{2 \pi i/8}$ and $\xi=e^{2 \pi i/9}$ and $a\in \Zp^*,\,b,c\in \Zp $.  
For a fixed value $a \in \Zp^*$ the magic states, in addition to the computational basis, form a complete set of MUBs since $|\braket{f_{a,b,c}}{f_{a,b^\prime,c^\prime}}|=\delta_{b,b^\prime}\delta_{c,c^\prime}+(1-\delta_{b,b^\prime})/\sqrt{p}$ \cite{Alltop:1980,Bengtsson:2014}.
Finally, define the \Jam state $\ket{J_{a,b,c}}$ as that which is created by applying a magic gate to one half of a Bell pair:
\begin{align}
&\ket{J_{a,b,c}}= \mathbb{I}_p \otimes M_{a,b,c} \ket{\Phi}, \quad\ket{\Phi}=\sum_j \frac{\ket{jj}}{\sqrt{p}}. \label{eqn:Jam}
\end{align}
Clearly $\ket{J_{a,b,c}}$ is isomorphic to a single-qudit magic state $\ket{f_{a,b,c}}$ under the identification $\ket{jj}\in \mathbb{C}^{p^2} \leftrightarrow \ket{j}\in \mathbb{C}^{p}$.

The group of operations generated by (tensor products of) the Weyl-Heisenberg operators in Eq.~\eqref{eqn:DispOps} is the Pauli group (we denote it $\mathcal{C}_1$ for reasons that will become clear). The set of unitaries that map the Pauli group into itself under conjugation is the Clifford group ($\mathcal{C}_2$). The unitaries that map the Pauli group into the Clifford group, under conjugation, define \cite{Gottesman:1999} the third level of the Clifford hierarchy, $\mathcal{C}_3$, and so on. Overall we have
\begin{align}
\mathcal{C}_k:=\{U|U \mathcal{C}_1 U^\dag \subseteq C_{k-1}\}
\end{align}
where $\mathcal{C}_1\subset \mathcal{C}_2 \subset \mathcal{C}_3 \ldots$ but the sets $\mathcal{C}_{k>3}$ no longer form a group. The operations in the first two levels arise naturally in the context of fault-tolerant quantum computation, whereas an operation from third level is typically also required to enable universal quantum computation. Properties of $\mathcal{C}_3$ for multiple qubits are discussed in \cite{Gottesman:1999} whereas the diagonal single-qudit subset of $\mathcal{C}_3$, as discussed in \cite{Howard:2012}, is given by $M_{a,b,c}$ above. It is in this sense that we say states $\ket{f_{a,b,c}}$ and $\ket{J_{a,b,c}}$ reflect the structure of (the third level of) the Clifford hierarchy.

%

\subsection{Generalizing the CHSH game}

The familiar two-qubit CHSH inequality takes the form $\langle \mathcal{B} \rangle_{\max}^{\textsc{LHV}} \leq 2$ with a Bell operator
\begin{align}
&\mathcal{B}=XX+XY+YX-YY=\sum_{x,y=0}^{1} \omega^{xy}A_xB_y\\
&\text{where } \quad \{A_0,A_1\}=\{B_0,B_1\}=\{X,Y\}
\end{align}
and $\omega=-1$ is a $p$-th root of unity. The maximum expectation value using local hidden variables (LHV) is $2$ whereas quantum mechanics can achieve $2\sqrt{2}$ which is the maximum eigenvalue, $\lambda_{\max}(\mathcal{B})$, of the Bell operator. The optimal shared state that Alice and Bob can measure in this case is given by $\ket{J_{1,0,0}}$ from Eq.~\eqref{eqn:Jam} i.e.,
\begin{align}
\langle \mathcal{B} \rangle_{\max}^{\text{QM}}=\bra{J_{1,0,0}}\mathcal{B}\ket{J_{1,0,0}}=2\sqrt{2}. \label{eqn:QubitOpt}
\end{align}
Generalizing the above CHSH Bell operator to qudits \cite{Ji:2008}, we find it convenient to define two closely related operators
\begin{align}
\mathcal{B}^*&=\frac{1}{p}\sum_{n\in\Zp^*} \sum_{x,y\in\Zp} \omega^{nxy}A_x^nB_y^n, \label{eqn:BellOpStar} \\ 
\mathcal{B}&=\frac{1}{p}\sum_{n,x,y\in\Zp} \omega^{nxy}A_x^nB_y^n. \label{eqn:BellOp}
\end{align}
where $\omega=e^{2\pi i/p}$ so that $\mathcal{B}=\mathcal{B}^*+p\mathbb{I}_{p^2}$ and consequently $\lambda_{\max}(\mathcal{B})=\lambda_{\max}(\mathcal{B}^*)+p$. The traceless version $\mathcal{B}^*$ is more convenient to work with but the maximal value of $\mathcal{B}$ is a more useful quantity. 

To aid our analysis we use the Pauli measurement operators adopted in \cite{Ji:2008}
\begin{align}
A_x&=\omega^{x(2x+1)/2}D_{(1|x)},\quad 
B_y=\omega^{y(y+1)/4}D_{(1|y/2)} \label{eqn:AxBy}
\end{align}
where $x,y \in \mathbb{Z}_p$, so that $A_{y/2}=B_y$. The operators $\{D_{(1|j)},j\in \Zp\}$ are a generalization of the qubit measurements $\{X,Y\}$ used above insofar as they are the Pauli observables (excluding $Z$) whose eigenbases form mutually unbiased bases. In the qudit case, different orderings and phases of the measurement operators can have an effect on the quantumly achievable expectation value. 

The expectation value of $\mathcal{B}$ from Eq.~\eqref{eqn:BellOp} can be related to the quantum value, $0\leq \nu \leq 1$, of a nonlocal game \cite{Brunner:2014,Buhrman:2010}, which takes the general form
\begin{align}
\nu=\sum_{a,b,x,y \in \Zp}P(x,y)V(a,b,x,y)P(a,b|x,y)
\end{align}
for measurement settings $(x,y)$, outcomes $(a,b)$ and a predicate or pay-off function $V\in\{0,1\}$. By choosing a flat probability distribution over all measurement settings, $P(x,y)=1/p^2$, and a winning condition $V(a,b,x,y)=\delta_{a +b +xy,0}$ we arrive at
\begin{align}
\nu=\frac{1}{p^2}\bra{\Psi} \mathcal{B}\ket{\Psi} 
 =\frac{1}{p^2}\sum_{\begin{smallmatrix}
a+b+xy=0
\end{smallmatrix}}P(a,b|x,y),\label{eqn:valCHSH}
\end{align}
where the right hand side  of the above expression is a more general form for CHSH-type games, amenable to POVMs as well as projective measurements (see also a discussion in \cite{Liang:2009}).

The maximum classical value $\langle \mathcal{B} \rangle_{\max}^{\text{LHV}}$ is given by finding local hidden variable assignments
\begin{align*}
\vec{a}&=(a_0,\ldots,a_x,\ldots,a_{p-1})\in \Zp^p\\
\vec{b}&=(b_0,\ldots,b_y,\ldots,b_{p-1})\in \Zp^p
\end{align*}
that tell Alice and Bob which element of the spectrum of their observable they should output for a given measurement setting i.e.,
\begin{align*}
A_x \mapsto \omega^{a_x} \quad , \quad B_y \mapsto \omega^{b_y}.
\end{align*}
The best choices of $\vec{a}$ and $\vec{b}$ maximize the quantity
\begin{align*}
\langle \mathcal{B} \rangle_{\max}^{\text{LHV}}&=\frac{1}{p}\sum_{n,x,y=0}^{p-1} \omega^{nxy}(\omega^{a_x})^n(\omega^{b_y})^n\\
&=\sum_{a,b,x,y \in \Zp} \delta_{a_x+b_y+xy,0}
\end{align*}
i.e., they give the best classical strategy for participants trying to maximize the number of instances where $a+b+xy=0 \bmod p$. Liang \emph{et al.} \cite{Liang:2009} provide an explicit classical strategy achieving
\begin{align*}
\langle \mathcal{B} \rangle_{\max}^{\text{LHV}}\geq 3p-2.
\end{align*}

Turning to quantum strategies, Bavarian and Shor's result \cite{Bavarian:2013} (restricted to $q=p$) says that the quantum value \eqref{eqn:valCHSH} of a generalized CHSH game is bounded above (via Information Causality \cite{Pawlowski:2009}) by
\begin{align}
\nu\leq IC(p):=\frac{1}{p}\left(1+\frac{p-1}{\sqrt{p}}\right) \label{eqn:IC}
\end{align}
Clearly this implies $\langle \mathcal{B} \rangle_{\max}^{\text{QM}}\leq p^2IC(p)$ but it also applies to more general scenarios wherein Alice and Bob use POVMs. It is also shown \cite{Bavarian:2013} that classical strategies can achieve $\Omega \left(\frac{1}{\sqrt{q}}\right)$ if $q=p^{2k}$ or $O\left(\frac{1}{\sqrt{q}q^{\epsilon_0}}\right)$ if $q=p^{2k-1}$ for a very small  ($\leq \frac{1}{700}$) positive constant $\epsilon_0$. The moral is that we might not expect a classical-quantum separation in the asymptotic limit. 
Liang \emph{et al.} \cite{Liang:2009} used the Navascues-Pironio-Acin \cite{Navascues:2007} hierarchy to numerically bound the quantum value $\nu$ from above and it appears that the values so obtained coincide with $IC(p)$ for $p=5,7$.

\subsection{Entropic Uncertainty Relations and Eavesdropping }\label{sec:EURs}
Fix the notation for the mutually unbiased basis (MUB) expansion of a generic operator $K$ acting on a qutrit state space as follows:
\begin{align*}
K\leftrightarrow\left(
\begin{array}{c|c|c|c}
 c_{0,\infty} & c_{0,0} & c_{0,1} & c_{0,2} \\
 c_{1,\infty} & c_{1,0} & c_{1,1} & c_{1,2} \\
 c_{2,\infty} & c_{2,0} & c_{2,1} & c_{2,2} \\
\end{array}
\right) 
\end{align*}
where 
\begin{align*}
c_{V,B}=\bra{\psi_B^V}K\ket{\psi_B^V}\qquad (\ket{\psi_B^V}\text{ from Eq.~\eqref{eqn:ketBV}})
\end{align*}
and the infinity basis label corresponds to the computational basis i.e., $\ket{\psi_{\infty}^V}=\ket{V}$. The generalization to higher dimensions is obvious. The coefficients $c_{V,B}$ correspond to probabilities when $K$ is a normalized density operator, and the probability distribution associated with a basis $B$ is given by $\{c_{0,B},c_{1,B},\ldots,c_{p-1,B}\}$. Inverting this decomposition, we find \cite{Ivonovic:1981}
\begin{align*}
K&=\sum_{B,V} c_{V,B} \ketbra{\psi_B^V}{\psi_B^V}-\Tr(K)\mathbb{I}_p.
\end{align*}

With the above decomposition in hand we can characterize a density matrix $\rho$ in terms of its associated probability distributions. The Renyi entropy of order 2 (i.e., the collision entropy) with respect to a particular basis $B$ is given by $H_2^B(\rho)=-\log \left(\sum_V c_{V,B}^2 \right)$ whereas the min-entropy is given by $H^B_{\min}(\rho)=-\log\ \max_V c_{V,B}$ \cite{Wootters:2007,Mandayam:2010}. For an arbitrary pure state $\ket{\phi}\in\mathbb{C}^p$ this can be rewritten as
\begin{align*}
H_2^B(\ket{\phi})&=-\log \left(\sum_V |\braket{\phi}{\psi_B^V}|^4 \right),\\
H^B_{\min}(\ket{\phi})&=-\log\ \max_V |\braket{\phi}{\psi_B^V}|^2,
\end{align*}
which obey
\begin{align*}
\log(p)\geq H_2^B(\ket{\phi}) \geq H_{\min}^B(\ket{\phi})\geq 0.
\end{align*}
Typically we are interested in total or average entropies across a number of different measurement bases. The trade-offs that arise when we try to minimize entropy across multiple measurement bases gives insights into the structure of quantum state space.

For a complete set of $p+1$ unbiased measurement bases, the total collision entropy associated with any pure state is bounded from below by \cite{Wootters:2007,Appleby:2014b}
\begin{align}
\sum_{B\in\{\infty,\Zp\}}-\log \sum_V |\braket{\phi}{\psi_B^V}|^4  &\geq-(p+1)\log \frac{2}{p+1} \label{eqn:MinEntBound}
\end{align}
The factor of $2$ in the numerator arises from the surprising fact that  $\sum_{B,V} |\braket{\phi}{\psi_B^V}|^4=2$ for \emph{any} pure state $\ket{\phi}$. The above inequality is saturated whenever $\sum_V |\braket{\phi}{\psi_B^V}|^4$ is independent of the basis $B$, as occurs for SIC-POVM fiducial states \cite{Appleby:2014b}. For this reason fiducial states qualify as ``minimum-uncertainty states'' (not all minimum uncertainty states are fiducial vectors, however). MUB-balanced states \cite{Amburg:2014} have the same probability distribution for all $p+1$ bases and consequently also saturate this lower bound.

Now, instead of collision entropy, consider the total min-entropy across a number of unbiased measurement bases. The following min-entropic uncertainty relation was proven for dimensions of the form $d=2^n$ in \cite{Mandayam:2010} and generally in \cite{Rastegin:2013}
\begin{align*}
\sum_{B\in\{\infty,\Zd\}} H^B_{\min}(\ket{\phi})&\geq -(d+1)\log \left[\frac{1}{d}\left(1+\frac{d-1}{\sqrt{d+1}}\right)\right].
\end{align*}
Later we will interested in the case there we omit the computational basis and restrict to prime-dimensional systems so the relevant lower bound becomes
\begin{align}
\sum_{B\in \Zp} H^B_{\min}(\ket{\phi})&\geq -p\log \left[\frac{1}{p}\left(1+\frac{p-1}{\sqrt{p}}\right)\right]. \label{eqn:HinfLB}
\end{align}

\begin{figure}
\begin{center}
\includegraphics[trim = 50mm 170mm 80mm 30mm,clip=true,width=0.2\textwidth]{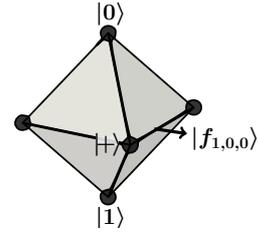}
\end{center}
\caption{\label{fig:Breidbart}%
 The Breidbart basis $\{\ket{f_{1,0,0}},\ket{f_{1,0,1}}\}$  (which is equal to $\{\ket{H},\ket{H^\perp}\}\}$ in the magic state notation of Bravyi and Kitaev \cite{Bravyi:2005}) constitutes the optimal measurement basis for Eve if her goal is to infer as much information about Alice's state as possible. The cryptographic protocol \cite{Bennett:1992,BB:1984,Williamson:2011} assumes Alice is sending one of the four equatorial Pauli eigenstates.}
\end{figure}

As well as constraining quantum state space, the above entropic quantities have an operational meaning in the context of cryptography \cite{Wehner:2010,Wootters:2007,Renes:2010,Wu:2009}. As explained in e.g., \cite{Williamson:2011}, if the four possible signal states transmitted from Alice to Bob in the BB'84 quantum key distribution protocol \cite{BB:1984} are the eigenstates of $\sigma_x$ and $\sigma_y$, then the optimal measurement basis (see Fig.~\ref{fig:Breidbart})  for Eve's intercept-resend attack is the basis of magic states $\{\ket{f_{1,0,0}},\ket{f_{1,0,1}}\}$ better known as $\{\ket{H},\ket{H^\perp}\}$ in the context of quantum computation or the Breidbart basis in cryptographic settings \cite{Bennett:1992,BB:1984,Williamson:2011}.
This basis, which can be thought of as splitting the difference between the $\sigma_x$ and $\sigma_y$ bases allows Eve to ascertain Alice's information with probability $\frac{1}{2}(1+\frac{1}{\sqrt{2}})\approx 85\%$, (which is the optimal value of the CHSH nonlocal game) and the state $\ket{H^{(\perp)}}$ minimizes the average min-entropy for the $X$ and $Z$ basis measurements \cite{Maassen:1988,Wehner:2010}. Finding higher-dimensional analogues of the Breidbart basis motivated Amburg et al. \cite{Amburg:2014} to investigate MUB-balanced states. We shall see how qudit magic states obey many of the same desirable properties of the Breidbart basis.

\section{Results}
Our results are broken up into two subsections, the first of which concerns nonlocality while the second concerns entropic uncertainty. In the first subsection, we find a simple expression for the eigenbasis of the CHSH-Bell operator \eqref{eqn:BellOp} which, after the application of a result from number theory, allows us to place good bounds on the quantum value of the restricted (to Pauli measurements) CHSH game. In the second subsection, we examine magic states in terms of their ability to minimize the total (or average) entropy across a number of different measurement bases. For this purpose we use collision entropy and min-entropy, both of which have operational significance in cryptographic settings. Among the subset of qudit states that we call equatorial states, magic states always minimize the collision entropy, and for small prime dimension they also simultaneously minimize the min-entropy.

\subsection{Optimal strategy for CHSH game and bounds on the achievable value}
Rewrite the Bell operator $\mathcal{B}^*$ of Eq.~\eqref{eqn:BellOpStar} in the Weyl-Heisenberg basis using Eq.~\eqref{eqn:AxBy}
\begin{align}
\mathcal{B}^*
&=\frac{1}{p}\sum_{n\in\Zp^*,x,s\in \Zp}(\omega^{s(s+\tfrac{1}{2})}D_{(1,1|x,s-x)})^n, 
\end{align}
and note that conjugation by a Clifford gate $C^\dag$ has the effect
\begin{align*}
C^\dag D_{(1,1|x,s-x)}^n C = D_{(1,0|s,s-x)}^n,
\end{align*}
where
\begin{align*}
C:=CSUM_{12}=\sum_k \ketbra{k}{k} \otimes X^{k} : \ket{j,k}\mapsto \ket{j,k+j}
\end{align*}
is the generalized CNOT (Controlled-NOT) operator.
Under this unitary operation the Bell operator becomes
\begin{align}
C^\dag \mathcal{B}^*C &=\frac{1}{p}\sum_{n\in\Zp^*,x,s\in\Zp}(\omega^{s(s+\tfrac{1}{2})}D_{(1,0|s,s-x)})^n\\
&=p\mathcal{S}^*\otimes \ketbra{0}{0}
\end{align}
with
\begin{align}
\mathcal{S}^*&=\frac{1}{p}\sum_{n\in\Zp^*,s\in\Zp} (\omega^{s(s+\tfrac{1}{2})}D_{(1|s)})^n\\
\text{and } \quad \mathcal{S}&=\frac{1}{p}\sum_{n,s\in\Zp} (\omega^{s(s+\tfrac{1}{2})}D_{(1|s)})^n
\end{align}
so that $\mathcal{S}=\mathcal{S}^*+p\mathbb{I}_p$ and $\lambda_{\max}(\mathcal{S})=\lambda_{\max}(\mathcal{S}^*)+p$ as before. By the definition of stabilizer projectors in Eq.~(\ref{eqn:projBV}) we have
\begin{align}
 \mathcal{S}&=\sum_{B} \ketbra{\psi_B^{-B(B+\frac{1}{2})}}{\psi_B^{-B(B+\frac{1}{2})}} \label{eqn:SasSum}
\end{align}
i.e. $\mathcal{S}$ is a sum of projectors, one from each of the non-computational bases.
The final step is to note that $\lambda_{\max}(\mathcal{B}^*)=\lambda_{\max}(p\mathcal{S}^*\otimes \ketbra{0}{0})=\lambda_{\max}(p\mathcal{S}^*)$ so that
\begin{align}
\lambda_{\max}(\mathcal{B})=p\lambda_{\max}(\mathcal{S}).
\end{align}

Overall, we find that the maximizing eigenvector of the Bell operator $\mathcal{B}$ gives an expectation that is the same as maximizing a simple single-qudit operator $\mathcal{S}$. Moreover, $\mathcal{S}$ is a sum of $p$ stabilizer projectors, one from each of the non-computational stabilizer bases. We will show that $\mathcal{S}$ is maximized by a magic state $\ket{f_{a,b,c}}$ as in Eq.~\eqref{eqn:fabcdef}. It follows that the optimum quantum strategy for our restricted CHSH game is for Alice and Bob to measure a shared state $\ket{J_{a,b,c}}$ of Eq.~\eqref{eqn:Jam} since
\begin{align*}
\bra{J_{a,b,c}} \mathcal{B}^* \ket{J_{a,b,c}}
= &\bra{f_{a,b,c},0}  C^\dag\mathcal{B}^*C \ket{f_{a,b,c},0}\\
= &p\bra{f_{a,b,c},0}  \left( \mathcal{S}^*\otimes \ketbra{0}{0}\right)\ket{f_{a,b,c},0}\\
= &p\bra{f_{a,b,c}} \mathcal{S}^*\ket{f_{a,b,c}}\\
=&p\lambda_{\max}(\mathcal{S}^*)
\end{align*}

\begin{table}[!h]
\begin{align*}
\begin{array}{c|cccc}
p & p(1+\frac{p-1}{\sqrt{p}})&  4p & \langle \mathcal{B} \rangle_{\max}^{\text{QM}}  & \langle \mathcal{B} \rangle_{\max}^{\text{LHV}} \\ \hline
 3 & 6.4641 & -  & 6.4115   & 6  \\
 5 & 13.9443& 20.  & 13.0902  & 12  \\
 7 & 22.8745& 28.  & 19.4112  & 19  \\
 11& 44.1662&  44. & 34.6464  & 37 \\
 13& 56.2666&  52. & 48.3481  & 47 \\
 17& 82.9697&  68. & 55.1022  & \geq 66 \\
 19& 97.4602&  76. & 72.6084  & \geq 79 \\
 23& 128.508&  92. & 74.8954  & \geq 99 \\
 29& 179.785&  116. & 104.819  & \geq 135\\
\end{array}
\end{align*}
\caption{\label{tab:comparison} Comparison of the bounds imposed by Information Causality (first column) and Weil (second column) with the optimal quantum and local hidden variable strategies (third and fourth columns). The LHV lower bounds are reproduced from \cite{Liang:2009} whereas the operator norm $\langle \mathcal{B} \rangle_{\max}^{\text{QM}} $ is attained by measuring the maximally entangled states $\ket{J_{a,b,c}}$. Because we have restricted ourselves to Pauli measurements, the quantity $\langle \mathcal{B} \rangle_{\max}^{\text{QM}} $ is less than what is achievable in the unrestricted case. } 
\end{table}

In Sec.~\ref{sec:ThmProof} we prove that the magic state $\ket{f_{a,b,c}}$, with $a=-1/12$ and $b=-1/8$ is a maximizing eigenvector for the single-qudit operator $\mathcal{S}$ for $p>3$. 
Moreover, we show that all magic states are balanced with respect to Alice and Bob's measurements projectors given in Eq.~\eqref{eqn:projBV} i.e.,
\begin{align}
\{|\braket{\psi_B^{V}}{f_{a,b,c}}|,V\in\Zp\} \quad \text{independent of basis } B.
\end{align}
Our proof provides an explicit expression showing exactly how probabilities $|\braket{\psi_B^{V}}{f_{a,b,c}}|^2$ are permuted between different bases. For the magic state $\ket{f_{-1/12,-1/8,c}}$ under consideration, the overlap with vector $V_B$ in basis $B$ is constant whenever
\begin{align}
V_B=-B\left(B+\frac{1}{2}\right).
\end{align}
Combining this with Eq.~\eqref{eqn:SasSum} we find
\begin{align}
\lambda_{\max} (\mathcal{S})&=\sum_{B} |\braket{\psi_B^{-B(B+\frac{1}{2})}}{f_{-1/12,-1/8,c}}|^2\\
&= p|\braket{+}{f_{-1/12,-1/8,c}}|^2 \label{eqn:lambdamax}
\end{align}
All overlaps between (non-computational) MUB vectors and magic states take the same form viz.
\begin{align}
|\langle \psi_B^V \ket{f_{a,b,c}}|^2&=\frac{1}{p}|\sum_k\omega^{ak^3+(b-2^{-1}B)k^2+(c-V)k}|^2
\end{align}
The values of these overlaps are constrained by the Weil bound \cite{Weil:1948,Bombieri:1966}, which says that a polynomial $f$ of degree $n$ over a prime field $\Zp$ satisfies
\begin{align}
|\sum_x\omega^{f(x)}|\leq (n-1)\sqrt{p}
\end{align}
provided $p$ does not divide $n$. Applied to Eq.~\eqref{eqn:lambdamax} the above bound implies
\begin{align}
\langle \mathcal{B} \rangle_{\max}^{\text{QM}}=\lambda_{\max} (\mathcal{S})\leq 4p. \label{eqn:WeilBound}
\end{align}
For $p>7$ this bound is more restrictive than the Information Causality bound. In Table \ref{tab:comparison} we compare this bound with $\lambda_{max}(\mathcal{S})=\langle \mathcal{B} \rangle_{\max}^{\text{QM}}$ and with the bound $p\left(1+\frac{p-1}{\sqrt{p}}\right)$ imposed by Information Causality. 

The qutrit case must be handled individually -- terms like $-1/12$ are ill-defined and the Weil bound is not applicable. One can explicitly check that the optimal quantum state for Alice and Bob to measure is what we have come to expect i.e. a maximally entangled state that is isomorphic to a magic state,
\begin{align}
\langle \mathcal{B} \rangle_{\max}^{\text{QM}}=\bra{J_{1,1,0}}\mathcal{B}\ket{J_{1,1,0}}=3\sqrt{3}\cos(\pi/18)=6.4115.
\end{align}
Liang \emph{et al.} \cite{Liang:2009} have shown that this is truly the quantum maximum; even allowing for POVMs this value cannot be surpassed.


\subsection{Minimum uncertainty equatorial states}

\begin{figure}[ht!]
\begin{center}
\includegraphics[width=0.5\textwidth]{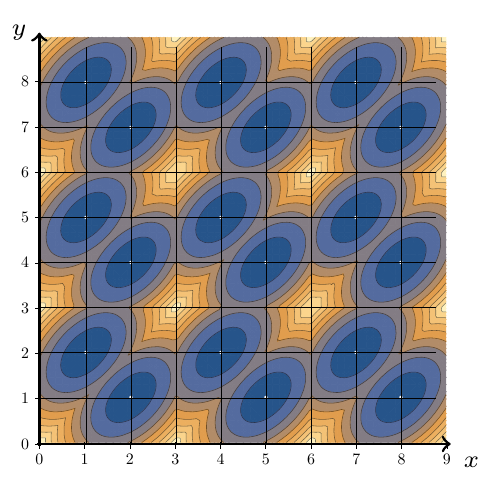}
\end{center}
\caption{\label{fig:Uncertainty}%
Total min-entropy across $3$ different bases: This plot shows contours of the total uncertainty for equatorial qutrit states parameterized as $\ket{\psi_{eq} (\vec{\phi})}=(1,\xi^x,\xi^y)/\sqrt{3}$ with $0\leq x,y \leq 9 \in \mathbb{R}$ and $\xi=e^{2 \pi i/9}$. The minimum uncertainty states occur at the small white dots, which are integer values of $x,y$ such that $x=2a+6b+3c \bmod 9$ and $y=a+6b+6c \bmod 9 $ with $a\in\mathbb{Z}_3^*$ and $b,c\in \mathbb{Z}_3$. In other words, magic states $\ket{f_{a,b,c}}$ as defined in Eq.~\eqref{eqn:fabc23} are minimum uncertainty states amongst the set of equatorial states. This optimality of magic states $\ket{f_{a,b,c}}$ holds numerically for prime dimensions $p\in\{2,3,5,7\}$ but not for $p=11$ or $p=13 $.}
\end{figure}
Define a generalized equatorial state in terms of a vector of phases $\vec{\phi}=(\phi_0:=0,\phi_1,\ldots,\phi_{p-1})$, 
\begin{align}
\ket{\psi_{eq} (\vec{\phi})}=\frac{1}{\sqrt{p}}\sum_{k=0}^{p-1} e^{i \phi_k}\ket{k}. \label{eqn:Equator}
\end{align}
This state has $p-1$ free real parameters, which is exactly half the number of parameters that specify an arbitrary pure state. This is already a very important class of states for cryptographic protocols e.g., optimal cloning procedures are known for states of this form \cite{Fan:2003,Scarani:2005}.

The total collision-entropic uncertainty across $p$ non-computational Pauli bases is bounded from below by
\begin{align}
\sum_{B\in\Zp}-\log \sum_V |\braket{\psi_{eq} (\vec{\phi})}{\psi_B^V}|^4  &\geq-p\log \left(\frac{2-\frac{1}{p}}{p}\right) 
\end{align}
since $\sum_V |\braket{\psi_{eq} (\vec{\phi})}{V}|^4=1/p$ and we already know from Eq.~\eqref{eqn:MinEntBound} that the sum over all $p+1$ bases is identically 2 for any pure state. Because of the balancedness property of $\ket{f_{a,b,c}}$ proven in Sec.~\ref{sec:ThmProof} we have that $\sum_{V}|\braket{f_{a,b,c}}{\psi_B^V}|^4$ is independent of $B$ and therefore the above inequality is saturated. In summary, of all equatorial states \eqref{eqn:Equator} it turns out that magic states minimize the total collision entropy and saturate the above entropic uncertainty relation.


%

Using magic states, we will not generally be able to saturate the lower bound of Eq.~\eqref{eqn:HinfLB} on the total min-entropy across $p$ mutually unbiased measurements. However, for small prime dimensions, it is feasible to check numerically whether
magic states minimize the total min-entropy when we restrict to the manifold of equatorial states defined in Eq.~\eqref{eqn:Equator}. To that end define
\begin{align}
\vec{\phi}_{\min}=&\argmin_{\vec{\phi}} \sum_{B \in \Zp} H^B_{\min} \left(\ket{\psi_{eq} (\vec{\phi})}\right).
\end{align}
For qutrits, see Fig.~\ref{fig:Uncertainty}, we find that
\begin{align}
\vec{\phi}_{\min}&= (0,\phi_1,\phi_2)\\
&=\frac{2\pi}{9}\left(0,2a+6b+3c,a+6b+6c\right) \nonumber 
\end{align}
whereas for primes $p=5$ or $p=7$ we find
\begin{align}
\vec{\phi}_{\min}=(\ldots,\phi_k,\ldots)=\frac{2\pi}{p}\left(ak^3+bk^2+ck\right) . 
\end{align}
This set of phases is exactly what defines the magic states $\ket{f_{a,b,c}}$ in \eqref{eqn:fabcdef} and \eqref{eqn:fabc23}
In some cases the total min-entropy they achieve is quite close to the bound that applies to generic states. e.g,
\begin{align}
p=3:\qquad \sum_{B \in \Zp} H^B_{\min} \ket{\psi_{eq} (\vec{\phi}_{\min})}=1.468
\end{align}
(where we used $\log_2$) which is not much greater than the lower bound
\begin{align}
-3\log_2 \left[\frac{1}{3}\left(1+\frac{2}{\sqrt{3}}\right)\right]=1.4324
\end{align}
Similarly for $p=5$ we get $4.667>4.2113$ and for $p=7$ (see Table \ref{tab:qusept}) we get $7.871>7.693$. For $p=11$ we find a counterexample to the optimality of $\ket{f_{a,b,c}}$ i.e., we find $\sum_{B \in \Zp} H^B_{\min} \ket{f_{a,b,c}}=19.8465>15.994$ but there is another non-magic equatorial state that achieves $17.7606$.  For $p=13$ a magic state of the form $\ket{f_{4,b,c}}$  achieves $23.471 > 20.6267$ but this can also be beaten by a non magic equatorial state that achieves a total min-entropy of $23.1336$.

Minimum uncertainty states are defined relative to a particular notion of entropy and we have chosen two instances that appear to have both operational and geometrical interpretations. Besides the collision entropy $H_2$ and the min-entropy $H_{\alpha=\infty}$, the general Renyi entropy $H_{\alpha}$ is well-defined for a one parameter family $\alpha \geq 0$, so it could be the case that magic states also minimize uncertainty for other values of $\alpha$. Minimum uncertainty states can also be defined relative to the Shannon entropy (see e.g. \cite{Coles:2011}) but the results so obtained are qualitatively very different.

\section{Theorems \& Proofs}\label{sec:ThmProof}

This section contains the two key mathematical results and their proofs. Many of their implications have already been used in preceding sections. The definitions and notation were established in Sec.~\ref{sec:Prerequisites}

\begin{theorem}
The magic state $\ket{f_{a,b,c}}$, with $a=-1/12$ and $b=-1/8$ is a maximizing eigenvector for the single-qudit operator $\mathcal{S}$ \eqref{eqn:SasSum} for all prime dimensions $p>3$.
\end{theorem}

We claim that the eigenbasis that diagonalizes $\mathcal{S}$ is given by the set of orthogonal magic states $\{\ket{f_{-1/12,-1/8,c}},c \in \Zp \}$. The proof is obtained by constructing a unitary $U$ with these states as columns and showing that conjugating $\mathcal{S}$ with $U^\dag$ produces a diagonal matrix $D$.
\begin{align*}
\text{Claim:}\quad D=U^\dag \mathcal{S} U= \left(\begin{array}{ccc}
\lambda_1(\mathcal{S}) & 0 & 0 \\ 
0 & \ddots & 0 \\ 
0 & 0 & \lambda_p(\mathcal{S})
\end{array}  \right)
\end{align*}
where $\lambda_i(\mathcal{S})$ are the eigenvalues in no particular order. Perform matrix multiplication to find the explicit matrix elements $d_{j,k}$ of our putative diagonal matrix, $D$, i.e.,
\begin{align*}
&D=\sum_{j,k}d_{j,k}\ketbra{j}{k}=U^\dag \mathcal{S} U=\\
&\left(\sum_{g,h} \omega^{\frac{g^3}{12}+\frac{g^2}{8}+gh} \ketbra{g}{h}\right) \mathcal{S} \left( \sum_{l,m} \omega^{-(\frac{m^3}{12}+\frac{m^2}{8}+lm)}\ketbra{l}{m} \right)
\end{align*}
Next, insert 
\begin{align*}
\mathcal{S}=s_{u,v} \ketbra{u}{v}=\sum_{B} \omega^{\frac{B}{2}(u^2-v^2)+B(B+\frac{1}{2})(u-v)}\ketbra{u}{v},
\end{align*}
which is a consequence of Eq.~(\ref{eqn:SasSum}) and Eq.~(\ref{eqn:ketBV}). After some tidying and relabelling we find
\begin{align*}
&d_{j,k}=\\
&\sum_{A,B,C \in \Zp}\omega^{\frac{1}{12}(A^3-C^3)+(\frac{B}{2}+\frac{1}{8})(A^2-C^2)+(\frac{B}{2}+B^2)(A-C)+kC-jA}
\end{align*}
Perform the linear substitution
\begin{align*}
&(A,B,C)\mapsto(\tilde{A}+\tilde{C},\tilde{B}-\tilde{A}/2-1/4,\tilde{A}-\tilde{C})\\
\Rightarrow\ &d_{j,k}=\sum_{\tilde{A},\tilde{B},\tilde{C} \in \Zp}\omega^{\frac{\tilde{C}^3}{6}+\tilde{C}(-\frac{1}{8}+2\tilde{B}^2-j-k)+\tilde{A}(k-j)}
\end{align*}
which satisfies
\begin{align*}
&d_{j,k\neq j}=0
\end{align*}
because
\begin{align*}
\sum_{\tilde{A} \in \Zp}\omega^{\tilde{A}(k-j)}=p\delta_{k,j} 
\end{align*}
Hence $D$ is diagonal, as claimed.
$\blacksquare$

\begin{theorem}
All magic states $\ket{f_{a,b,c}}$ are MUB-balanced with respect to the mutually unbiased measurements $\{X,XZ,\ldots,XZ^{p-1}\}$. Moreover, the $V_B$-th coefficient in the $B$-th basis is given by the $V_0$-th coefficient in the $X$ measurement basis when $V_B=V_0+\frac{1}{12a}\left(B^2-4Bb\right)$
\end{theorem}

First define the quantity $T_{a,b,c}$ as
\begin{align*}
T_{a,b,c}=\frac{1}{p}|\sum_k \omega^{ak^3+bk^2+ck}|
\end{align*}
and note that, after a linear substitution $k \mapsto k+r$ (where $r$ is some fixed element $r \in \Zp$) that does not change the value of the sum, we find
\begin{align}
T_{a,b,c}=T_{a,b+3ar,c+2br+3ar^2} .\label{eqn:Tequivalence}
\end{align}

Take the probability $c_{V_0,0}$ associated with an arbitrary vector $V_0$ in the first measurement basis, $B=0$:
\begin{align}
c_{V_0,0}=& |\braket{\psi_0^{V_0}}{f_{a,b,c}}|^2 \nonumber \\
=&\frac{1}{p}|\sum_k \omega^{ak^3+bk^2+(c+V_0)k}|^2 \nonumber \\
=& T_{a,b,c+V_0} \label{eqn:T1}
\end{align}
Now consider a vector $V_B$ in any other basis $B$
\begin{align}
c_{V_B,B}=&|\braket{\psi_{B}^{V_B}}{f_{a,b,c}}|^2 \nonumber \\
=&\frac{1}{p}|\sum_k \omega^{ak^3+(b-B/2)k^2+(c+{V_B})k}|^2 \nonumber \\
=&T_{a,b-B/2,c+{V_B}} \label{eqn:T2}
\end{align}
Then, with the aid of Eq.~(\ref{eqn:Tequivalence}), we find the condition that makes Eq.~\eqref{eqn:T1} and Eq.~\eqref{eqn:T2} equal:
\begin{align}
c_{V_0,0}&=c_{V_B,B}, \qquad  \forall B \in \Zp\\
\Rightarrow V_B&=V_0+\frac{1}{12a}\left(B^2-4Bb\right) \qquad \blacksquare
\end{align}

A geometrical argument for the balancedness of magic states was implicit in recent work of Blanchfield \cite{Blanchfield:2014} (see the proof of Theorem 1 therein) although that study concerned relationships between different MUB constructions. Not only have we proven that magic states $\ket{f_{a,b,c}}$ are MUB-balanced, we have shown exactly which permutation of probabilities occurs when moving between bases.

This theorem applied to the magic state $\ket{f_{-1/12,-1/8,c}}$ that maximizes $\mathcal{S}$ gives
\begin{align}
V_B&=\frac{1}{12(-1/12)}\left(B^2-4B(-1/8)\right)=-B(B+\frac{1}{2})
\end{align}
which makes intuitive sense given that 
\begin{align}
\mathcal{S}&=\sum_{B} \ketbra{\psi_B^{-B(B+\frac{1}{2})}}{\psi_B^{-B(B+\frac{1}{2})}}.
\end{align}
Using the MUB decomposition introduced in the Sec.~\ref{sec:EURs}, the single-qutrit operator $\mathcal{S}$ looks like 
\begin{align*}
&\mathcal{S}=\ketbra{\psi_0^0}{\psi_0^0}+\ketbra{\psi_1^0}{\psi_1^0}+\ketbra{\psi_2^1}{\psi_2^1},\\
\Rightarrow\quad &\mathcal{S}\leftrightarrow\left(
\begin{array}{c|c|c|c}
 1 & 5/3 & 5/3 & 2/3  \\
 1 & 2/3 & 2/3 & 5/3 \\
 1 & 2/3 & 2/3 & 2/3\\
\end{array}
\right), 
\end{align*}
while the maximizing eigenstate takes the form
\begin{align*}
\ketbra{f_{1,1,0}}{f_{1,1,0}}\leftrightarrow\left(
\begin{array}{c|c|c|c}
 1/3 & 0.7124 & 0.7124 & 0.0859  \\
 1/3 & 0.2017 & 0.2017 & 0.7124 \\
 1/3 & 0.0859 & 0.0859 & 0.2017\\
\end{array}
\right). 
\end{align*}

\section{Symmetries, Cyclers and Sato-Tate}\label{sec:Symmetries}

In this section we outline some symmetries of the states and operators that we have previously used. We then mention a connection between our work and that of Amburg \emph{et al.} \cite{Amburg:2014} .

To begin with, there is an obvious \emph{asymmetry} associated with magic states $\ket{f_{a,b,c}}$; they treat the computational basis differently to the remaining Weyl-Heisenberg (Pauli) bases. There is nothing special about the computational $Z$ basis; in fact we can exhaustively create all all $p^2(p^2-1)$ single-qudit $\ket{f}$-type magic states by applying the Clifford unitary that maps $Z \mapsto X$ followed by (multiple applications) of the Clifford that maps $X \mapsto XZ$. Any of these magic states will be balanced with respect to all but one Pauli measurement bases. Conversely, we lose nothing by restricting our analysis to the equatorial class of magic states that we have so far considered.

The $b$ and $c$ components of magic states are modified by the action of Pauli operators,
\begin{align}
&\ket{f_{a,b,c}}&=D_{(x|z)} \ket{f_{a,0,0}}, \qquad \left(\begin{array}{c}
x \\ 
z
\end{array} \right)=\left(\begin{array}{c}
-\tfrac{b}{3a} \\ 
c-\tfrac{b^2}{3a}
\end{array} \right)
\end{align}
whereas changing the value of $a$ in $\ket{f_{a,b,c}}$ requires at least a Clifford operation. In fact, in \cite{Bengtsson:2014} it was shown that magic states $\ket{f_{a,b,c}}$ and $\ket{f_{a^\prime,b^\prime,c^\prime}}$  with different values $a\neq a^\prime$ are connected via a Clifford if and only if these values lie in the same equivalence class of cubic residues modulo $p$. The noticeable effect of this is that in prime dimensions $p=1 \bmod 3$ there are $3$ equivalence classes of magic state that have different operational and geometrical features. In Table \ref{tab:qusept} we examine seven-dimensional magic states in terms of different operationally relevant quantities and see interesting differences arising from the inequivalent classes.

One of the main approaches to finding minimum uncertainty states has been to find so-called MUB cyclers -- a single unitary that, when repeatedly applied to a basis state, maps basis vectors from one basis to the next, eventually wrapping around cyclically to return to the original basis. Eigenstates of these unitaries are often minimum-uncertainty states. These unitaries are known to not exist in prime dimensions, so the unitary we define below is the best possible alternative -- it cycles through all but one of the bases. Define a Clifford gate $C_{a,b,c}$ as follows
\begin{align}
C_{a,b,c}=M_{a,b,c}XM_{a,b,c}^\dag,
\end{align}
then it follows from the definitions that $C_{a,b,c}\ket{f_{a,b,c}}= \ket{f_{a,b,c}}$. Consider $\protect{C:=C_{-1/12,-1/8,c}}$, then
\begin{align}
C^r\ket{\psi_0^0}=\ket{\psi_{-\frac{r}{2}}^{V_r}}, \qquad V_r=\tfrac{r}{2}(-\tfrac{r}{2}+\tfrac{1}{2})
\end{align}
so that for $p=5$ we find that repeated application of $C$ takes us through the basis indices $(0,2,4,1,3,0,2,4,\ldots)$.
It is fairly straightforward to prove the equivalent claim
\begin{align}
C\ket{\psi_{-\frac{s}{2}}^{V_s}}=\ket{\psi_{-\frac{s+1}{2}}^{V_{s+1}}}\qquad \forall s \in \Zp,
\end{align}
using the following argument,
\begin{align}
MXM^\dag\ket{\psi_{-\frac{s}{2}}^{V_s}}&=MX\ket{f_{\frac{1}{12},\frac{1-2s}{8},-(c+V_S)}},\\
&=M\sum_k\omega^{\frac{(k-1)^3}{12}+\frac{(k-1)^2(1-2s)}{8}-(k-1)(c+V_s)}\ket{k}\nonumber,\\
&=M\ket{f_{\frac{1}{12},-\frac{(2s+1)}{8},\frac{s}{2}-(c+V_S)}},\\
&=\ket{f_{0,-\frac{s+1}{4},\frac{s}{2}-V_S}},\\
&=\ket{\psi_{-\frac{s+1}{2}}^{V_s-\frac{s}{2}}}=\ket{\psi_{-\frac{s+1}{2}}^{V_{s+1}}}.
\end{align}
Hence, magic states are eigenstates of MUB-cycling Clifford unitaries that cycle through $p$ out of $p+1$ bases.

For prime dimensions $p>3$ magic states (as defined in Eq.~\eqref{eqn:fabcdef}) are equal weighted superpositions of roots of unity with a cubic polynomial in the exponent. Exponential sums are commonplace in number theory and many results are known. The  distribution
\begin{align}
\theta_{a,c}=\frac{\sum_k \omega^{ak^3+ck}}{2\sqrt{p}} \quad a\in \Zp^*,\,c\in \Zp\label{eqn:realcubic}
\end{align}
is real and lies in the range $[-1,1]$ \cite{Livne:1987,Browning:2010}.
 The behaviour of Eq.~\eqref{eqn:realcubic} for different values of $a,c$ and $p$ is covered by the Sato-Tate conjecture (see \cite{Browning:2010} Sec.~1.4). Hence, magic states $\ket{f_{a,0,c}}$ obey the same semicircular distribution of values $-\frac{2}{\sqrt{p}}\leq \braket{+}{f_{a,0,c}} \leq \frac{2}{\sqrt{p}}$ observed by Amburg \emph{et al.} \cite{Amburg:2014,Appleby:2014,Katz:2012} for their states $\ket{\beta}$ that are balanced with respect to all $p+1$ mutually unbiased bases. Also, $\theta_{a,c}$ becomes increasingly equidistributed  as $p$ grows to infinity \cite{Livne:1987} meaning that the Weil bound Eq.~\eqref{eqn:WeilBound} is saturated in this limit.

\begin{table}
\begin{align*}
\begin{array}{c|cccc}
a & W_{\min} &  Mana & \sum_B H^B_{\min}  &\\ \hline
 1& -0.027692 & 0.814835 &7.87055    \\
 2& -0.089915 & 0.814835 &12.3287  \\
 3& -0.034531 & 0.896212 &9.35125   \\
 4& -0.034531 & 0.896212 &9.35125  \\
 5& -0.089915 & 0.814835 &12.3287  \\
 6& -0.027692 & 0.814835 &7.87055  \\
\end{array}
\end{align*}
\caption{\label{tab:qusept} Different characterizations of seven-dimensional magic states : Magic states have surprising additional structure in prime dimensions of the form $p=1 \bmod 3$ \cite{Bengtsson:2014}, which opens the door for operational inequivalence of magic states $\ket{f_{a,b,c}}$ depending on the value of the parameter $a\in\{1,2,\ldots,p-1\}$. The nonzero elements of $\mathbb{Z}_7$ can be partitioned into equivalence classes of cubic residues and their cosets i.e., $\mathbb{Z}_7^*=\{1,6\}\cup\{2,5\}\cup\{3,4\}$. Depending on which of the operationally relevant quantities ($W_{\min}, Mana$ or $\sum_B H^B_{\min}$) we focus on, we find a different choice of optimal equivalence class. The quantity $W_{\min}$ denotes the minimal value of the Wigner function of $\ket{f_{a,b,c}}$ in phase space \cite{Veitch:2012}, while $Mana$ effectively describes the magnitude of the sum of all the negative entries \cite{Veitch:2014}. The last column describes the total min-entropy of $\ket{f_{a,b,c}}$ across all measurement bases $\{X,XZ,\ldots,XZ^{p-1}\}$. Note that lower bound for arbitrary (not necessarily equatorial) states is $7.693=-7\log[(1+(7-1)/\sqrt{7})/7]$ so the magic state performs well.}
\end{table}

\section{Discussion and Conclusions}

Characterizing highly non-stabilizer states and their operational capabilities is important. Stabilizer states appear naturally within the context of fault-tolerant quantum computation and also as basis vectors for complete sets of mutually unbiased bases. 
 Here we have picked out a natural family of highly non-stabilizer states and showed their optimality in terms of both nonlocality and minimizing entropic uncertainty, where we have restricted to Pauli measurements in both scenarios. Finding concise optimal solutions to problems like these is of independent interest.

Whenever Alice and Bob are restricted to using qudit Pauli measurements we cannot ever observe nonlocality by measuring a stabilizer state. This is due to the existence of a local, noncontextual hidden variable model for non-negatively represented states in a particular discrete Wigner function \cite{Gross:2006,Veitch:2012,Howard:2013}. By showing that magic states give the maximal violation relative to a stabilizer CHSH scenario, this gives an operational interpretation to the concept of being highly non-stabilizer. Similarly,  a stabilizer state is a maximally uncertain state with respect to a set of mutually unbiased Pauli bases. By showing that magic states minimize the entropic uncertainty, this gives another operational characterization of their highly non-stabilizer character. 

One could argue the case for a number of different ways of quantifying non-stabilizerness. The correct metric will most likely depend on the exact nature of the task for which this state or operation is a resource.
In \cite{Veitch:2014} two measures were put forward that were tailored toward fault-tolerant quantum computation via magic state distillation.
The relative entropy between the test state and the closest positively represented state, as well as the sum-negativity (the sum of all the negative quasi-probabilities in the discrete Wigner function) of the state were both highlighted. It is worth noting the disagreement in Table \ref{tab:qusept} between the state $\ket{f_{3,b,c}}$ that Mana picks out as opposed to the state $\ket{f_{1,b,c}}$ that min-entropy picks out. Other studies have characterized non-stabilizer states and operations in terms of convex geometry  \cite{WvDMH:2011} and in terms of frame potentials \cite{Andersson:2014}.

For qubits, the Tsirelson bound for the CHSH scenario can be saturated using Pauli measurements applied to a particular maximally entangled state. The maximally nonlocal state $\ket{J_{1,0,0}}$ \eqref{eqn:QubitOpt} can be understood as the image of a ``pi-over-eight'' gate applied to one half of a Bell pair $\ket{\Phi}$. It is intuitively pleasing that this connection between optimality and the Clifford hierarchy continues for all higher prime values of $p$, as we have shown here. Less pleasing is that the quantum advantage diminishes with increasing dimension, and seems to disappear entirely for $p\geq 11$ (we highlight as an interesting open question whether \emph{any} alternate choice of Pauli measurements \eqref{eqn:AxBy} for Alice and Bob could result in $\lambda_{\max}(\mathcal{B})>\langle \mathcal{B} \rangle_{\max}^{\text{LHV}}=37$). For $p<11$ it should be borne in mind that \cite{Jungnitsch:2010} showed a clear operational advantage for restricting to stabilizer measurements when statistical significance, rather than the amount of Bell inequality violation, is adopted as the figure of merit. One possible practical application of our results is the benchmarking of non-stabilizer resources for fault-tolerant computation. Any imperfect $\ket{J_{a,b,c}}$ (or equivalently $M_{a,b,c}$) that still allows for violation of the qudit Bell-CHSH inequality is also suitable for promoting fault-tolerant Clifford gates to universal quantum computation via magic state distillation \cite{Howard:2012b}. This is a sufficient condition but not necessary, as the distillation routines of \cite{Campbell:2014} can also enable universality with imperfect $\ket{J_{a,b,c}}$ that have lost the ability to violate the Bell inequalities \cite{Ji:2008} that we used here.


The inter-relationship between entanglement, nonlocality, steering, complementarity and uncertainty is a fascinating and ongoing research program.
Oppenheim and Wehner \cite{Oppenheim:2010} have shown that the presence of uncertainty limits the amount of nonlocality, whereas Tomamichel and H\"anggi \cite{Tomamichel:2013} have shown that uncertainty is necessary in order to observe nonlocality.
In the qubit case, the similarity between the Information Causality bound of Eq.~\eqref{eqn:IC} and the entropic uncertainty bound of Eq.~\eqref{eqn:HinfLB} arises because of these links between nonlocality and uncertainty. The fact that the qudit expressions also have the same form suggests that this connection may hold generally, although this is complicated by the fact that the qudit CHSH game is no longer an XOR game \cite{Bavarian:2013}.

%
%


%


There are some obvious avenues for further investigation. Magic states $\ket{f_{a,b,c}}$ reflect the structure of the third level of the Clifford hierarchy, and take a simple exponential sum form for higher dimensions. Both of these facets are amenable to further analysis using group-theoretical and number-theoretical ideas. Moreover, they both suggest natural generalizations for classes of states that may prove useful, analagous to what we have proven here. The Clifford hierarchy is well-defined for multiple particles (e.g. controlled-Clifford gates are elements of the third level) but for single-particle diagonal gates the relationship between hierarchy level and the order of polynomials in the exponential sum is more transparent. The Weil bound \eqref{eqn:WeilBound} suggests that states defined as exponential sums of higher order polynomials may sometimes perform better. Finally, as with all calculations over prime fields, it is worth investigating how many of our results carry over to the case where we consider systems of prime-power dimensions $q=p^r$ by using the field $\Fq$ rather than $\Fp$.


\section{Acknowledgements}
We thank Jyrki Lahtonen for assistance regarding exponential sums, Earl Campbell for helpful suggestions and Joel Wallman for comments on an early draft. MH acknowledges financial support from FQXi, CIFAR and the Government of Canada through NSERC. This research was supported by the US Army Research Office through grant W911NF-14-1-0103.


\begin{thebibliography}{99}



 \bibitem{Acin:2002}
A. Ac\'in, T. Durt, N. Gisin, and J. I. Latorre
``Quantum nonlocality in two three-level systems''
\newblock \href{http://dx.doi.org/10.1103/PhysRevA.65.052325}{Phys. Rev. A 65, 052325 (2002).}

 \bibitem{Acin:2004}
A.~Ac\'in, J.~L.~Chen, N.~Gisin, D.~Kaszlikowski, L.~C.~Kwek, C.~H.~Oh, and M.~Zukowski,
``Coincidence Bell Inequality for Three Three-Dimensional Systems''
\newblock \href{http://dx.doi.org/10.1103/PhysRevLett.92.250404}{Phys.~Rev.~Lett. \textbf{92}, 250404, (2004).}

\bibitem{Durt:2003}
T.~Durt, N.~J.~Cerf, N.~Gisin, and M.~Zukowski
\newblock ``Security of quantum key distribution with entangled qutrits''
\newblock \href{http://dx.doi.org/10.1103/PhysRevA.67.012311}{Phys. Rev. A 67, 012311 (2003)}

\bibitem{CHSH:1969}
J.~F.~Clauser, M.~A.~Horne, A.~Shimony, and R.~A.~Holt
``Proposed Experiment to Test Local Hidden-Variable Theories''
\newblock \href{http://dx.doi.org/10.1103/PhysRevLett.23.880}{Phys. Rev. Lett. 23, 880 (1969).}


\bibitem{Brunner:2014}
N.~Brunner, D.~Cavalcanti, S.~Pironio, V.~Scarani, and S.~Wehner
``Bell nonlocality''
\newblock \href{http://dx.doi.org/10.1103/RevModPhys.86.419}{Rev. Mod. Phys. 86, 419  (2014)}


\bibitem{Buhrman:2010}
H.~Buhrman, R.~Cleve, S.~Massar, and R.~de Wolf
``Nonlocality and communication complexity''
\newblock \href{http://dx.doi.org/10.1103/RevModPhys.82.665}{Rev. Mod. Phys. 82, 665 (2010)}




\bibitem{Reichardt:2013}
B.~W.~Reichardt, F.~Unger and U.~Vazirani
``Classical command of quantum systems''
\newblock \href{http://dx.doi.org/10.1038/nature12035}{Nature 496, 456–460 (2013)}




\bibitem{Buhrman:2005}
H.~Buhrman and S.~Massar,
``Causality and Tsirelson's bounds''
\newblock \href{http://dx.doi.org/10.1103/PhysRevA.72.052103}{Phys. Rev. A 72, 052103 (2005).}


\bibitem{Ji:2008}
SW.~Ji, J.~Lee, J.~Lim, K.~Nagata, and HW.~Lee,
``Multisetting Bell inequality for qudits''
\newblock \href{http://dx.doi.org/10.1103/PhysRevA.78.052103}{Phys. Rev. A 78, 052103 (2008)}


\bibitem{Liang:2009}
YC.~Liang, CW.~Lim, and DL.~Deng,
``Reexamination of a multisetting Bell inequality for qudits''
\newblock \href{http://dx.doi.org/10.1103/PhysRevA.80.052116}{Phys. Rev. A 80 052116 (2009).}



\bibitem{Bavarian:2013}
M.~Bavarian, P.~W.~Shor,
``Information Causality, Szemerédi-Trotter and Algebraic Variants of CHSH''
\newblock \href{http://arxiv.org/abs/1311.5186}{arXiv:1311.5186 (2013).}



\bibitem{Pawlowski:2009}
M.~Paw{\l}owski, T.~Paterek, D.~Kaszlikowski, V.~Scarani, A.~Winter and M.~{\.Z}ukowski,
``Information causality as a physical principle''
\newblock \href{http://dx.doi.org/10.1038/nature08400}{Nature, \textbf{461} pp.~1101-1104 (2009).}


\bibitem{Jungnitsch:2010}
B.~Jungnitsch, S.~Niekamp, M.~Kleinmann, O.~G\"uhne, H.~Lu, W.~Gao, Y.~Chen, Z.~Chen, and J.~Pan
``Increasing the Statistical Significance of Entanglement Detection in Experiments''
\newblock \href{http://dx.doi.org/10.1103/PhysRevLett.104.210401}{Phys. Rev. Lett. 104, 210401 (2010)}

\bibitem{Acin:2005}
A.~Ac\'in, R.~Gill, and N.~Gisin
``Optimal Bell Tests Do Not Require Maximally Entangled States''
\newblock \href{http://dx.doi.org/10.1103/PhysRevLett.95.210402}{Phys. Phys. Rev. Lett. 95, 210402 (2005)}





\bibitem{Renes:2004} J.~M.~Renes, R.~Blume-Kohout, A.~J.~Scott, and C.~M.~Caves,
``Symmetric informationally complete quantum measurements'',
\newblock \href{http://dx.doi.org/10.1063/1.1737053}{ J.~Math.~Phys.~\textbf{45} 2171 (2004).} 


\bibitem{Gross:2006}
D.~Gross,
``Hudson’s theorem for finite-dimensional quantum systems''
\newblock \href{http://dx.doi.org/10.1063/1.2393152}{J.~Math,~Phys.~\textbf{12} 47 122107 (2006).}



\bibitem{Veitch:2012}
V.~Veitch, C.~Ferrie, D.~Gross and J.~Emerson,
``Negative quasi-probability as a resource for quantum computation''
\newblock \href{http://dx.doi.org/10.1088/1367-2630/14/11/113011}{New Journal of Physics \textbf{14},11 pp.~113011, (2012).}



\bibitem{Veitch:2014}
V.~Veitch, S.~A.~H.~Mousavian, D.~Gottesman and J.~Emerson
``The resource theory of stabilizer quantum computation''
\newblock \href{http://dx.doi.org/10.1088/1367-2630/16/1/013009}{ New J.~Phys.~\textbf{16} 013009 (2014)}



\bibitem{Howard:2014}
M.~Howard,	
J.~Wallman,	
V.~Veitch and
J.~Emerson,
``Contextuality supplies the ‘magic’ for quantum computation''
\newblock \href{http://dx.doi.org/10.1038/nature13460}{Nature, \textbf{510} pp.~351-355 (2014).}


\bibitem{Wehner:2010}
S.~Wehner and A.~Winter
``Entropic uncertainty relations -- a survey ''
\newblock \href{http://dx.doi.org/10.1088/1367-2630/12/2/025009}{New J. Phys. 12 025009 (2010).}


\bibitem{Amburg:2014}
I.~Amburg, R.~Sharma, D.~Sussman, W.~K.~Wootters
\newblock ``States that "look the same" with respect to every basis in a mutually unbiased set''
\newblock \href{http://arxiv.org/abs/1407.4074}{arXiv:1407.4074 [quant-ph]}




\bibitem{Appleby:2014}
D.~M.~Appleby, I.~Bengtsson, H.~B.~Dang
\newblock ``Galois Unitaries, Mutually Unbiased Bases, and MUB-balanced states''
\newblock \href{http://arxiv.org/abs/1409.7987}{arXiv:1409.7987 (2014)}



\bibitem{Wootters:2007}
W.~K.~Wootters, D.~M.~Sussman
``Discrete phase space and minimum-uncertainty states''
\newblock \href{http://arxiv.org/abs/0704.1277}{arXiv:0704.1277, (2007)}






\bibitem{Mandayam:2010}
P.~Mandayam, S.~Wehner and N.~Balachandran,
``A transform of complementary aspects with applications to entropic uncertainty relations''
\newblock \href{http://dx.doi.org/10.1063/1.3477319}{J. Math Phys \textbf{51}, 8, 082201 (2010).}



\bibitem{Campbell:2012}
E.~T.~Campbell, H.~Anwar and D.~E.~Browne
 ``Magic state distillation in all prime dimensions using quantum Reed-Muller codes'',
\newblock \href{http://link.aps.org/doi/10.1103/PhysRevX.2.041021}{Phys.~Rev.~X \textbf{2}, 041021, (2012).}


\bibitem{Campbell:2014}
E.~T.~Campbell,
 ``Enhanced Fault-Tolerant Quantum Computing in d-Level Systems'',
\newblock \href{http://dx.doi.org/10.1103/PhysRevLett.113.230501}{Phys.~Rev.~Lett \textbf{113}, 230501, (2014).}



 \bibitem{Howard:2012}
M.~Howard and J.~Vala,
``Qudit versions of the qubit $\pi/8$ gate''
\newblock  \href{http://dx.doi.org/10.1103/PhysRevA.86.022316}{Phys.~Rev.~A. \textbf{86}, 022316, (2012).}


\bibitem{Alltop:1980}
W.~O.~Alltop,
``Complex sequences with low periodic correlations''
\newblock \href{http://dx.doi.org/10.1109/TIT.1980.1056185}{IEEE Trans. Inform. Theory \textbf{26} 350 1980}


\bibitem{Bengtsson:2014}
I.~Bengtsson, K.~Blanchfield, E.~Campbell and M.~Howard,
``Order 3 Symmetry in the Clifford Hierarchy''
\newblock \href{http://dx.doi.org/10.1088/1751-8113/47/45/455302}{J.~Phys.~A: Math.~Theor. \textbf{47} 455302 (2014)}




\bibitem{Gottesman:1999} 
D.~Gottesman and I.~L.~Chuang,
``Demonstrating the viability of universal quantum computation using teleportation and single-qubit operations'',
\newblock \href{http://dx.doi.org/10.1038/46503}{Nature \textbf{402}, 390 (1999)}.


\bibitem{Navascues:2007}
M.~Navascues ,S.~Pironio and A.~Ac\'in ,
``Bounding the Set of Quantum Correlations''
\newblock \href{http://link.aps.org/doi/10.1103/PhysRevLett.98.010401}{Phys. Rev. Lett. \textbf{98} 010401 (2007).}



 \bibitem{Ivonovic:1981}
I.~D.~Ivonovic
``Geometrical description of quantal state determination''
\newblock \href{http://dx.doi.org/10.1088/0305-4470/14/12/019}{J.~Phys.~A. \textbf{14}, 3241, (1981)}.



\bibitem{Appleby:2014b}
D.~M.~Appleby , H.~B.~Dang  and C.~A.~Fuchs
``Symmetric Informationally-Complete Quantum States as Analogues to Orthonormal Bases and Minimum-Uncertainty States''
\newblock \href{http://dx.doi.org/10.3390/e16031484}{Entropy, \textbf{16}, 3, pp.~1484-1492 (2014)}

\bibitem{Rastegin:2013}
A.~E.~Rastegin,
``Uncertainty relations for MUBs and SIC-POVMs in terms of generalized entropies''
\newblock \href{http://dx.doi.org/10.1140/epjd/e2013-40453-2}{Eur. Phys. J. D  \textbf{67} 269 (2013).}

\bibitem{Renes:2010}
J.~M.~Renes
``Duality of privacy amplification against quantum adversaries and data compression with quantum side information ''
\newblock \href{http://dx.doi.org/10.1098/rspa.2010.0445}{Proceedings of the Royal Society A 467, 1604 (2011).}

\bibitem{Wu:2009}
S.~Wu, S.~Yu and K.~M\o lmer
``Entropic uncertainty relation for mutually unbiased bases''
\newblock \href{http://dx.doi.org/10.1103/PhysRevA.79.022104}{Phys. Rev. A 79, 022104 (2009).}


\bibitem{Williamson:2011}
M.~Williamson, V.~Vedral
\newblock ``Eavesdropping on practical quantum cryptography''
\newblock \href{http://www.tandfonline.com/doi/abs/10.1080/09500340308235253}{J. Mod. Opt. 50(13), 1989-2011 (2003)}


\bibitem{BB:1984}
C.~H.~Bennett, G.~Brassard,
``Quantum cryptography: Public key distribution and coin tossing''
\newblock \href{http://www.cs.ucsb.edu/~chong/290N-W06/BB84.pdf}{Proceedings of IEEE International Conference on Computers, Systems and Signal Processing, pp. 175-179 (1984)}

\bibitem{Bennett:1992}
C.~H.~Bennett, F.~Bessette, G.~Brassard,L.~Salvail and J.~Smolin, 
``Experimental quantum cryptography''
\newblock \href{http://dx.doi.org/10.1007/BF00191318}{Journal of Cryptology, \textbf{5}, 1, 3--28 (1992) }


\bibitem{Maassen:1988}
H.~Maassen and J.~B.~M.~Uffink
``Generalized entropic uncertainty relations ''
\newblock \href{http://dx.doi.org/10.1103/PhysRevLett.60.1103}{Phys. Rev. Lett. 60, 1103 (1988).}

\bibitem{Weil:1948}
A.~Weil,
``On some exponential sums''
\newblock \href{http://www.pnas.org/content/34/5/204.full.pdf+html}{Proc. Nat. Acad. Sci. USA , \textbf{34} pp.204--207 (1948).} 


\bibitem{Bombieri:1966}
E.~Bombieri
``On Exponential Sums in Finite Fields''
\newblock \href{http://www.jstor.org/stable/2373048}{American Journal of Mathematics, \textbf{88}, 1, pp.~71--105 (1966)}


\bibitem{Fan:2003}
H.~Fan, H.~ Imai, K.~Matsumoto, and X.~Wang
``Phase-covariant quantum cloning of qudits''
\newblock \href{http://dx.doi.org/10.1103/PhysRevA.67.022317}{Phys. Rev. A 67, 022317 (2003)}


\bibitem{Scarani:2005}
V.~Scarani, S.~Iblisdir, N.~Gisin, and A.~Ac\'in
``Quantum cloning''
\newblock \href{http://dx.doi.org/10.1103/RevModPhys.77.1225}{Rev. Mod. Phys. 77, 1225 (2005)}


\bibitem{Coles:2011}
P.~J.~Coles, L.~Yu, M.~Zwolak
Relative entropy derivation of the uncertainty principle with quantum side information
\newblock \href{http://arxiv.org/abs/1105.4865}{arXiv:1105.4865 (2011).}


\bibitem{Blanchfield:2014}
K.~Blanchfield,
``Orbits of mutually unbiased bases''
\newblock \href{http://dx.doi.org/10.1088/1751-8113/47/13/135303}{J.~Phys.~A: Math.~Theor.~\textbf{47} 135303 (2014).}



\bibitem{Browning:2010}
T.~D.~Browning
``Exponential sums over finite fields'',
\newblock Lecture Notes (2010), available at
\href{http://www.maths.bris.ac.uk/~matdb/tcc/EXP/EXP.pdf}{http://www.maths.bris.ac.uk/~matdb/tcc/EXP/EXP.pdf}



\bibitem{Livne:1987}
R.~Livn\'e. 
``The average distribution of cubic exponential sums.''
\newblock \href{http://eudml.org/doc/152918}{ J. Reine Angew.
Math., \textbf{375/376} 362--379, (1987)}.


\bibitem{Katz:2012}
N.~M.~Katz
\newblock ``Rigid local systems and a question of Wootters''
\newblock \href{http://dx.doi.org/10.4310/CNTP.2012.v6.n2.a1}{Communications in Number Theory and Physics \textbf{6}, 223 (2012)}




\bibitem{Howard:2012b}
M.~Howard and J.~Vala
``Nonlocality as a benchmark for universal quantum computation in Ising anyon topological quantum computers''
\newblock \href{http://dx.doi.org/10.1103/PhysRevA.85.022304}{Phys. Rev. A 85, 022304 (2012).}



\bibitem{Howard:2013}
M.~Howard,	
E.~Brennan and
J.~Vala,
``Quantum Contextuality with Stabilizer States''
\newblock \href{http://dx.doi.org/10.3390/e15062340}{Entropy, 15(6), 2340-2362 (2013).}

\bibitem{Bravyi:2005} 
S.~Bravyi and A.~Kitaev
Universal quantum computation with ideal Clifford gates and noisy ancillas
\newblock \href{http://dx.doi.org/10.1103/PhysRevA.71.022316}{Phys. Rev. A 71, 022316 (2005).}



\bibitem{Oppenheim:2010}
J.~Oppenheim and S.~Wehner,
``The Uncertainty Principle Determines the Nonlocality of Quantum Mechanics''
\newblock \href{http://dx.doi.org/10.1126/science.1192065}{Science \textbf{330} 1072--1074 (2010)}


\bibitem{Tomamichel:2013}
M.~Tomamichel and E.~H\"anggi,
``The link between entropic uncertainty and nonlocality''
\newblock \href{http://dx.doi.org/10.1088/1751-8113/46/5/055301}{J.~Phys.~A \textbf{46}, 5, 055301 (2013).}


\bibitem{WvDMH:2011}
W.~van Dam and M.~Howard
``Noise thresholds for higher-dimensional systems using the discrete Wigner function''
\newblock \href{http://dx.doi.org/10.1103/PhysRevA.83.032310}{Phys. Rev. A 83, 032310 (2011)}


\bibitem{Andersson:2014}
D.~Andersson, I.~Bengtsson, K.~Blanchfield and H.~B.~Dang
``States that are far from being stabilizer states''
\newblock \href{http://arxiv.org/abs/1412.8181}{arXiv:1412.8181 (2014)}






%





















%










%




\end{thebibliography}
\end{document}